\def\aap{{\em Astronomy and Astrophysics}}

\def\araa{{\em Annual Review of Astronomy and Astrophysics}}
\def\apj{{\em The Astrophysical Journal}}

\def\apjl{{\em The Astrophysical Journal Letters}}

\def\grl{{\em Geophys. Res. Lett.}}
\def\jgr{{\em J. Geophys. Res.}}

\def\mpla{{\em Modern Physics Letters A}}

\def\npg{{\em Nonlinear Processes in Geophysics}}
\def\nu{{\em Nuclear Fusion}}
\def\prl{{\em Phys. Rev. Lett.}}

\def\pop{{\em Phys. Plasma}}
\def\pof{{\em Phys. Fluid}}

\def\physrep{{\em Physics Reports}}

\def\rmp{{\em Rev.~Mod.~Phys.}}

\def\ssr{{\em Space Sci.~Rev.}}
\def\solphys{{\em Solar Physics}}
\def\sjetp{{Soviet Journal of Experimental and Theoretical Physics}}

\documentclass{pnastwo}
\usepackage[numbers, square, comma, sort&compress]{natbib}
\url{www.pnas.org/cgi/doi/10.1073/pnas.0709640104}
\copyrightyear{2008}
\issuedate{Issue Date}
\volume{Volume}
\issuenumber{Issue Number}
\footlineauthor{Che et al.}

\begin{document}

\title{How Electron Two-Stream Instability Drives Cyclic Langmuir Collapse and Continuous Coherent Emission}

\author{H. Che \affil{1}{Department of Astronomy, University of Maryland, College Park, MD, 20742, USA} \affil{2}{Heliospheric Physics Laboratory, Goddard Space Flight Center, NASA, Greenbelt, MD, 20771, USA},
M. L. Goldstein\affil{2}{},
P. H. Diamond\affil{3}{Department of Physics, University of California, San Diego, La Jolla, CA, 92093, USA},
\and
R. Z. Sagdeev\affil{4}{Department of Physics, University of Maryland, College Park, MD, 20742, USA}}

\contributor{Direct submission to Proceedings of the National Academy of Sciences of the United States of America, published on 1/31/2017}

\significancetext{Electron two-stream instability (ETSI) is believed to be responsible for the radio bursts observed in both the solar corona and interplanetary medium. What nonlinear kinetic processes self-consistently reconcile the several orders of magnitude difference between the short linear saturation time scale of ETSI and the long duration of bursts is fundamental in plasma turbulence and remains unsolved for nearly 50 years. Using massive particle-in-cell simulations, we find the first self-consistent and complete nonlinear solution to this problem.}

\maketitle

\begin{article}
\begin{abstract}
{Continuous plasma coherent emission is maintained by repetitive Langmuir collapse driven by the nonlinear evolution of a strong electron two-stream instability. The Langmuir waves are modulated by solitary waves in the linear stage, and by electrostatic whistler waves in the nonlinear stage. Modulational instability leads to Langmuir collapse and electron heating that fills in cavitons. The high pressure is released via excitation of a short wavelength ion acoustic mode that is damped by electrons and that re-excites small-scale Langmuir waves---this process closes a feedback loop that maintains the continuous coherent emission. }
\end{abstract}

\abbreviations{ETSI, Electron Two-stream Instability; ES, Electrostatic; EM, Electromagnetic; IAW, ion acoustic wave; KAW, kinetic Alfv\'en wave; LC, Langmuir Collapse; PIC, Particle-in-Cell}

\section{Introduction}
Electron beams accelerated by solar flares and nanoflares are believed to be responsible for several types of solar radio bursts observed in the corona and interplanetary medium, including flare-associated coronal Type U and J and interplanetary Type III radio bursts, and nanoflare-associated weak coronal type III bursts\cite{wild63araa,saint02sol,aschw02ssr,saint13apj}. In 1958 Ginzburg \& Zhelezniakov first proposed a basic framework for such bursts, which was subsequently refined by others\cite[references therein]{ginzburg59book,robinson97rmp}. In essence, the scenario is one in which the electron two-stream instability (ETSI), driven by electron beams, generates Langmuir waves that are converted into plasma coherent emission via nonlinear three-wave coupling (e.g. two Langmuir waves and one ion acoustic wave (IAW)). However, the mechanism whereby the nonlinear ETSI produces coherent emission with a duration of several orders of magnitude longer than the linear saturation time is not well understood
\cite{papa74apj,smith79apj,goldman80pof,robinson97rmp}. 
Nonlinear evolution of ETSI is a fundamental problem in nonlinear wave theory in which disparate three-wave couplings dominate the energy transport and dissipation\cite{ sagdeev69book}. It has broad applications in plasma physics, planetary and astrophysics, such as terahertz emission in laser beam experiments, radio bursts from Jupiter, pulsars and the formation of the exotic astrophysical objects. 

In the classical Kolmogorov turbulence scenario, the balance between energy input and its final absorption is controlled by a nonlinear cascade from large spatial scales (the region of external forcing) to viscosity-dominated short wavelengths. In plasmas, the source of instability is often beams of charged particles that generate Langmuir waves. At shorter wavelengths, the natural candidate to provide the sink of wave energy is Landau damping. However, nonlinear disparate wave interactions, it follows from direct calculation of basic three-wave coupling, can only lead to inverse cascades (to longer wavelengths) through modulational instability\cite{rud78physrep}, and away from the Landau damping region of the spectrum. The eventual nonlinear process capable of overriding this inverse cascade was suggested by Zakharov, namely Langmuir collapse (LC), which is analogous to a self-focusing of the Langmuir waves packets, or cavitons\cite{zak72sjetp,papa78grl}. LC has been discovered in both experiments and space observations\cite{wong84prl,kellogg92grla,ergun08prl}, in particular, LC has been observed in association with solar radio bursts\cite{kellogg92grla,ergun08prl,theja13jgr}. 

There are several problems with the current models of Type III radio bursts. The most important physics missed is the feedback of LC that we will show plays a critical role in maintaining continuous coherent emission. Second, the existing models\cite[and references therein]{robinson97rmp} use a common assumption that we call ``weak turbulence condition", specifically, the growth rate of Landau fluctuation driven by an electron beam is much smaller than that of the ETSI, or $v_{te,c}<v_b<<(n_c/n_b)^{2/3} v_{te,b}$, where $v_b$ is the electron beam drift, $v_{te,c}$ and $v_{te,b}$ are the thermal velocity of core background electrons and beams, respectively, and $n_c/n_b$ is the core-beam density ratio. However, recent observations suggest that the electron beam density near coronal source regions is comparable to the background density\cite{dennis11ssr}. Thus, the weak turbulence condition can be significantly violated. Third, the models assume that the emission is produced by coupling between Langmuir waves and IAWs, but IAWs are expected to be heavily damped in the nearly isothermal plasma of the corona.

\begin{figure}
\includegraphics[scale=0.45,trim=170 100 150 115,clip]{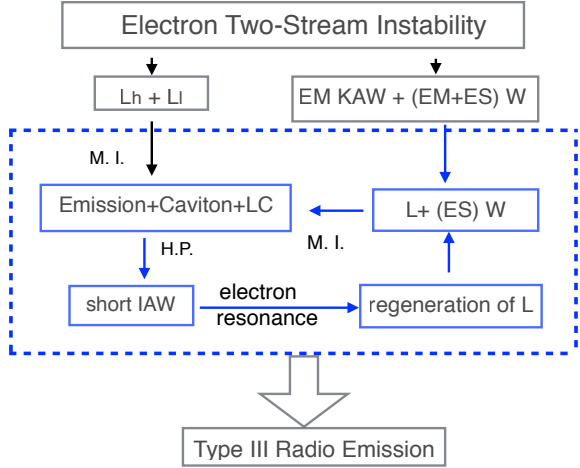}
\caption{Schematic diagram showing how LC, occurring during the nonlinear stage of the ETSI forms a feedback loop (within the blue-dashed line box) that produces coherent emission continuously. Meaning of the acronyms: M.I.: modulational instability, H.P.: high pressure, L: Langmuir wave, $L_h$: Langmuir wave with higher frequency produced by the background electrons, $L_l$: Langmuir wave with lower frequency produced by the trapped electrons in solitary waves, $W$: Whistler wave.  }
\label{chart}
\end{figure}

We here present a mechanism based on a model of cyclic LC and Langmuir wave regeneration. The results of massive particle-in-cell (PIC) simulations of the ETSI show how the nonlinear ETSI produces coherent emission that lasts 5 orders of magnitude longer than the linear saturation time. As shown in Fig.\ref{chart}, the extended emission time is a consequence of repeated LC, which regenerates Langmuir waves through resonance with intermediary short wavelength IAWs. The short wavelength IAW is produced due to the release of the ions inside the caviton caused by LC. Near the linear saturation of the ETSI, LC is initiated by the interactions between the high frequency Langmuir waves produced in the background and the low frequency Langmuir waves in the solitary wave trapped-electrons. As the ETSI enters the nonlinear decay, LC regenerates Langmuir waves and interacts with electrostatic (ES) whistler waves and re-initiate LC, thus forming a feedback loop. ES whistler waves are sustained by electromagnetic (EM) kinetic Alfv\'en waves (KAWs) and whistler waves that are produced simultaneously with the Langmuir waves. 

The structure of this paper is as follows: we first present the simulation results on the generation and regeneration of LC and emission during the nonlinear stage of ETSI. We then give the governing equations and condition for LC. Finally we show how LC regenerates Langmuir waves.
%

\section{Simulation Results of Cyclic Emission}
\label{intr}
\begin{figure}
\includegraphics[scale=0.45, angle=0,trim=30 260 60 200,clip]{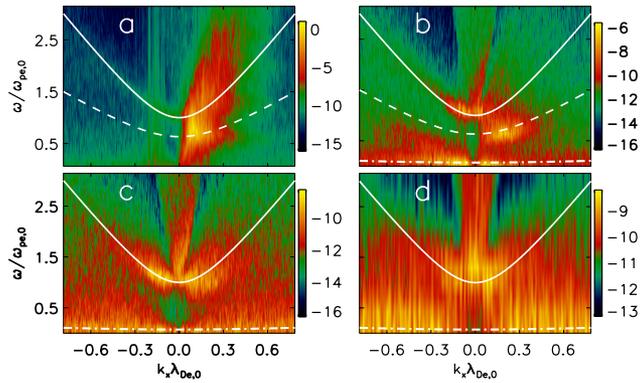} 
\caption{The  $\omega/\omega_{pe,0}-k_x \lambda_{De,0}$ diagrams of the parallel propagating high frequency electric field component $E_x$ at four time intervals:  (a): $\omega_{pe,0}t=0-100$; (b): $\omega_{pe,0}t=320-420$; (c): $\omega_{pe,0}t=2880-2980$; and  (d): $\omega_{pe,0}t=10560-10580$. Also shown are the dispersion relations (Eq.~\ref{lw1}) of the background electrons (solid lines), trapped electrons (dashed lines), and the short wavelength IAW (dash-dotted lines). The waves in space see supplementary information. The 2.5D PIC simulation is initialized with a homogeneous plasma and uniform magnetic field $\mathbf{B}=B_0 \hat{x}$. The initial ion velocity distribution function is a single Maxwellian and electron's is a core-beam bi-Maxwellian\citep{che14prl}. The initial density ratio of the beam and core is $n_{b0}/n_{c0} = 0.1$ and the core beam temperatures $T_{b,0}=T_{c,0}$. The initial drifts of the core $v_{cd,0}$ and the beam $v_{bd,0}$ satisfy $(1-\delta)v_{cd,0}= -\delta v_{bd,0}$ to maintain null current. $v_{bd,0}=12 v_{te,0} = 60 v_{A,0}$, where $v_{A0}$ is the initial Alfv\'en speed and $v_{te,0}=(k T_{c,0}/m_e)^{1/2}$. The speed of light $c = 100 v_{A,0}$ and $m_i/m_e=100$. The ion temperature $T_{i,0} = T_{c,0}$. The boundaries are periodic and the box size $L_x = L_y =3200 \lambda_{De,0}$ ($\lambda_{De,0}\equiv v_{te,0}/\omega_{pe,0}$), where $\omega_{pe,0}$ is the initial electron plasma frequency. The electric field is normalized by $E_0=v_{A,0} B_0/c$. $\beta=16\pi kT_{c,0}/B_0^2=0.25$.}
\label{expar}
\end{figure}

The initialization of 2.5D PIC simulation is described in the caption of Fig.\ref{expar}. The ratio between the initial beam velocity and the thermal velocity $ v_{bd,0}/ v_{te,0}\sim 12>(n_c/n_b)^{2/3}\sim 5$, where $v_{te,0}$ is the initial thermal velocity of both core and beam electrons, guarantees a strong ETSI. The total simulation time is $\omega_{pe,0} t=14400$ during which ETSI experiences a linear and nonlinear stage, saturation and nonlinear decay, and eventually reaches turbulent equilibrium, where the energy exchange between particles and waves reaches balance\cite{che14prl}.

\begin{figure}
\includegraphics[scale=0.45, angle=0,trim=30 260 60 180,clip]{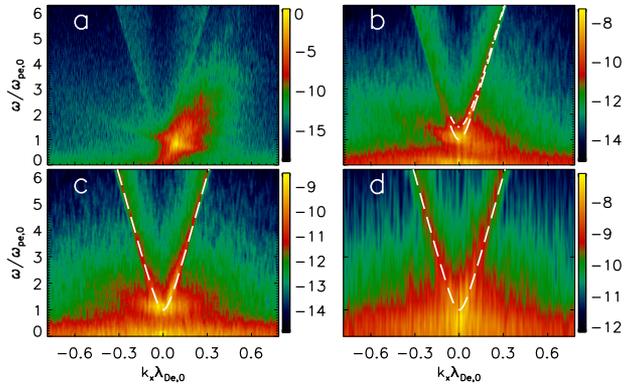} 
\caption{The  $\omega/\omega_{pe,0}-k_x \lambda_{De,0}$ diagrams of parallel propagating high frequency $E_y$ are shown for the same four time intervals as for Fig.\ref{expar}. Dashed lines: dispersion relation of plasma emission with frequency $\sim \omega_{pe,0}$. Solid line in (b): dispersion relation of plasma emission with frequency $\omega/\omega_{pe,0}=1.6$.} 
\label{eypar}
\end{figure}

The growth stage of ETSI includes the linear and nonlinear stage ($\omega_{pe,0}t=0$ to 200). The saturation stage is from $\omega_{pe,0} t\sim 200$ to 1000. The linear stage only lasts for $\omega_{pe,0} t=2$ as defined by the growth rate of the ETSI from quasi-linear theory (i.e.,$\gamma=\sqrt{3}/2(n_{b0}/n_{c0})^{1/3}\omega_{pe,0}\sim 2\omega_{pe,0}$). During the growth stage, the large beam drift suppresses the generation of Langmuir waves\cite{che16mpla}. The fastest growing mode of the solitary wave has $k_x \lambda_{De,0}\sim v_{te,0} /v_{bd,0}\sim 0.1$ and $\omega/\omega_{pe,0}\sim (n_{b0}/n_{c0})^{1/3}\omega_{pe,0}\sim 0.5 \omega_{pe,0}$ as shown in panels (a) in Fig.\ref{expar} \& ~\ref{eypar}. Quickly, the ETSI loses $\sim$ 85\% kinetic energy of the beams and reaches saturation with the beam drift $v_{db}$ being about two times of the core thermal velocity $\sim 2v_{te,c}\sim 30 v_{A,0}$. The thermal velocity of the electron beams $v_{te,b}$ increases to 40 $v_A$ and a bump forms at the tail of the core electron velocity distribution function\cite{che14apjl}. The core-beam density ratio changes to $n_b/n_c\sim 0.05$. The ratio $ v_{bd}/v_{te,b}\sim 0.7<(n_c/n_b)^{2/3}\sim 3$ indicates that the ETSI becomes weak turbulence. The bump starts to excite Langmuir waves as well as coherent emission (panels (b) in Fig.\ref{expar} \& ~\ref{eypar}).

The backward propagating Langmuir waves with frequency near $\omega_{pe,0}$ are excited in the background plasma while the propagating forward Langmuir waves with frequency near $0.6 \omega_{pe,0}$ are excited by the trapped electrons due to the low density and high temperature in the electron potential well\cite{stix92book}. These two Langmuir waves satisfy the following dispersion relation (normalized by the initial $\omega_{pe,0}$ and $\lambda_{De,0}$):
\begin{equation}
\frac{\omega}{\omega_{pe,0}}=\left ( \frac{n_e^2}{n_0^2}+\frac{T_{ce} n_e}{T_{ce,0} n_0} \gamma k_x^2\lambda_{De,0}^2\right )^{1/2},
\label{lw1}
\end{equation}
where $\gamma = 3$ as the electron heating caused by the solitary wave is nearly adiabatic \cite{che13pop}. 

The coalescence of the two anti-parallel Langmuir waves drives modulational instability and leads to LC\cite{zak72sjetp,rud78physrep}, accompanied by a 
harmonic emission with $\sim 1.6 \omega_{pe,0}$ (see Appendix ). The emission is shown in Fig.\ref{eypar}~(b), propagating much stronger forward than backward and satisfying the dispersion relation:
 \begin{equation}
\frac{\omega}{\omega_{pe,0}}=\left (\frac{n_e^2}{n_0^2} +\frac{c^2}{v_{te}^2}k_x^2\lambda_{De,0}^2\right )^{1/2}.
\label{light}
\end{equation}

\begin{figure}
\includegraphics[scale=0.55,trim=100 0 160 0,clip]{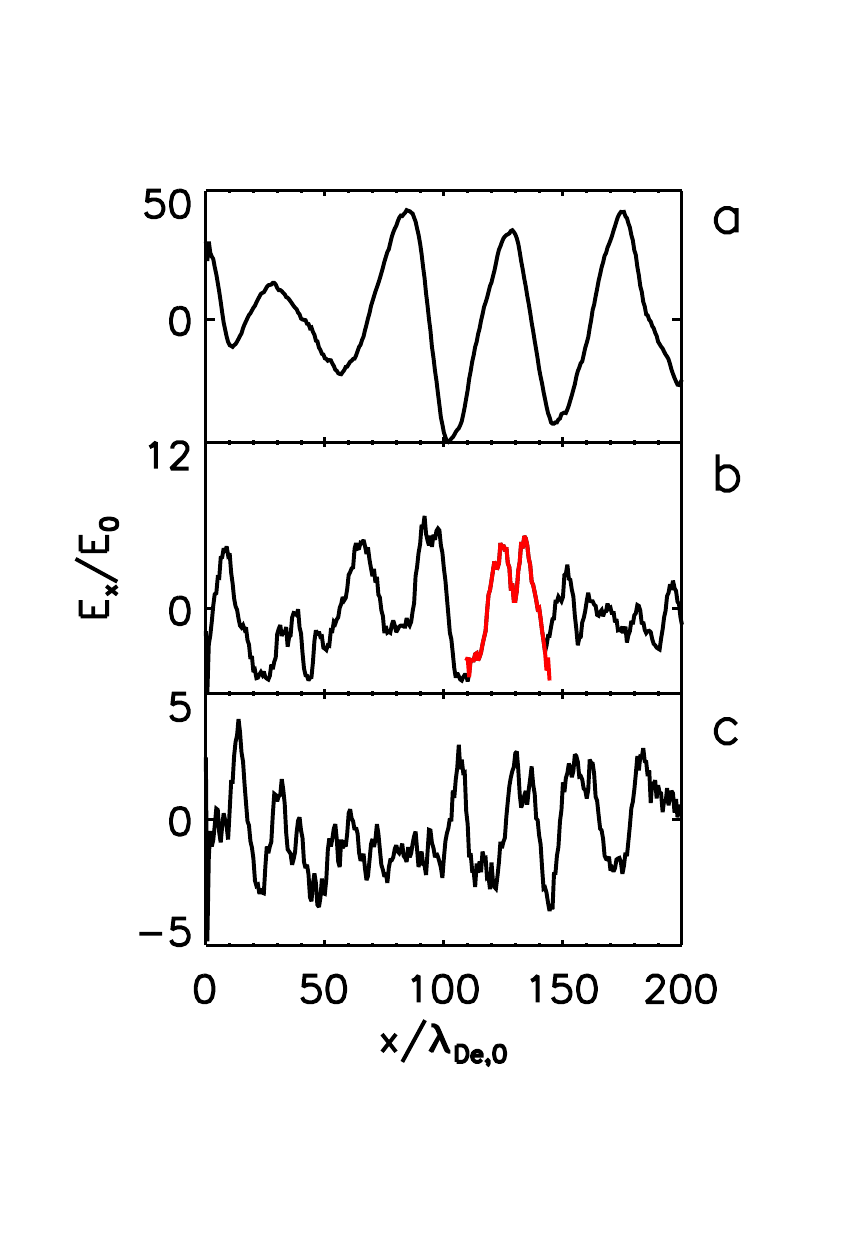} 
\includegraphics[scale=0.5,trim=50 200 0 220,clip]{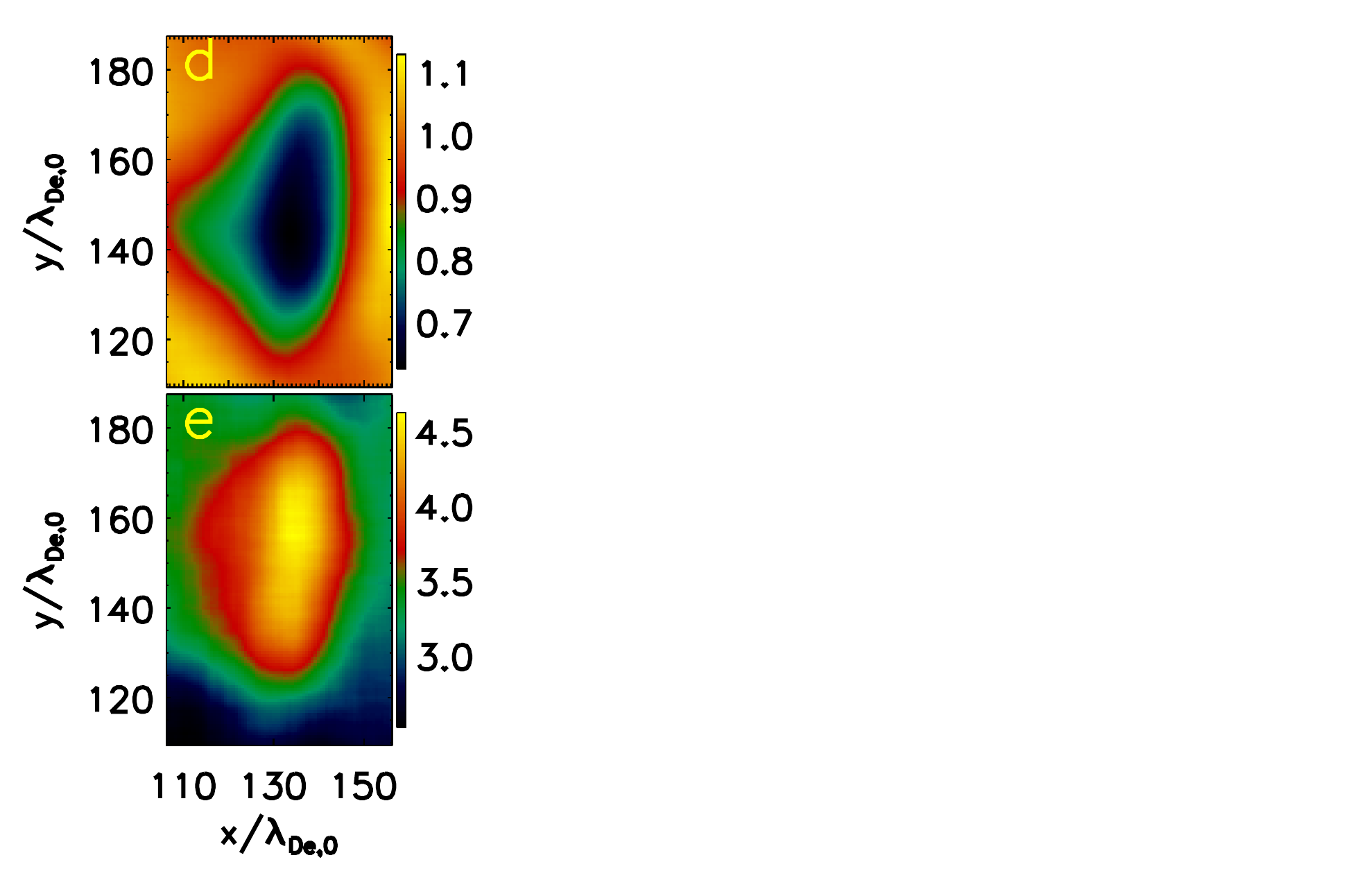} 
\caption{\textbf{Left panel} is the parallel electric field $E_x$ cuts for $x\in [0, 200] \lambda_{De,0}$ and $y=100 \lambda_{De,0}$ at a) $\omega_{pe,0} t=72$ when the ETSI nearly saturates and hot electrons excites Langmuir waves; b)$\omega_{pe,0} t=320$ the modulational instability grows, LCs start and cavitons form; c)  $\omega_{pe,0} t=648$ LCs continue. \textbf{Right panel}: the ion density in $xy$-plane i.e. the caviton (panel d) and the electron temperature in x direction $T_{ex}$ inside the caviton (panel e) corresponding to the Langmuir envelope in red in panel (b)(see supplementary movie I).  }
\label{collapse}
\end{figure}

The LC leads to the contraction of the modulated Langmuir envelope and the formation of ion density cavitons (see supplementary movie I). We plot a sample of parallel electric field $E_x$ in Fig.\ref{collapse} (a, b, c) at three moments: $\omega_{pe,0} t=$ 72, 320, 680. At $\omega_{pe,0} t=$ 72, the solitary waves with wavelength near the fastest growing mode reach the peak. The critical condition for LC $E^2/8\pi n_0 T_e>\frac{1}{4}k_x^2\lambda_{De}^2$ is satisfied since $(E^2/8\pi n_0 T_e)^{1/2}\sim 0.4$ with $E/E_0\sim 50$ is larger than the fastest growing mode of the ETSI  $k\lambda_{De}/2\sim v_{te,0}/2v_{db,0}\sim 0.05$. At $\omega_{pe,0} t=$ 320, the modulated wave envelopes decrease from 50 to 30~$\lambda_{De,0}$ and ion density cavitons form. In 
Fig.\ref{collapse}(d), we show an example of caviton in the $xy$-plane for the Langmuir envelope plotted in red in Fig.\ref{collapse}(b)(ref. supplementary movie I). Contraction of the Langmuir wave envelopes efficiently dissipates the Langmuir wave energy into electron thermal energy since the rate of Landau damping is proportional to $n_b \omega_{pe}^3/n_0 v_{te,b}^2 k^2$\cite{che16mpla}. The electron temperature along the magnetic field $T_{ex}$ inside the caviton is shown in panel (e). The increased pressure inside the caviton releases the excess density and produces an intermediary short IAW with frequency $\omega_{pi}$ (dash-dotted line in Fig.\ref{expar} b). The time scale for the growth of caviton is consistent with the modulation instability growth rate $\omega_{pi}(\langle E^2\rangle/8\pi n_0 T_e)^{1/2} \sim  0.01\omega_{pe}$. At $\omega_{pe,0} t=680$, some wave envelopes continue to contract to wavelengths about 10~$\lambda_{De,0}$ while some collapses lead to the vanishing of cavitons. 

\begin{figure}
\includegraphics[scale=0.6,trim=30 210 60 180,clip]{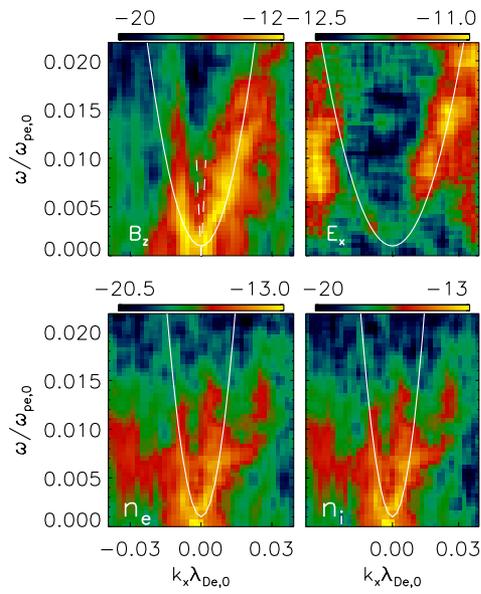}
\caption{The $\omega/\omega_{pe,0}-k_x \lambda_{De,0}$ diagram of parallel propagating low frequency fluctuations of $B_z$ in panel (a), $E_x$ in panel (b), density $n_e$ in panel (c) and density $n_i$ in panel (d). The data is from $\omega_{pe,0} t=$ 720 to 14400. Solid line: the dispersion relation of whistler wave; Dashed line: the dispersion relation of parallel propagating KAW. The details on the generation of KAWs and whistler waves by the nonlinear ETSI can be found in \cite{che14prl}. }
\label{disp}
\end{figure}

With the onset of LC, ETSI enters the nonlinear decay stage at $\omega_{pe,0}t \sim 1000$. The LC in saturation stage causes the two anti-propagating Langmuir waves to merge into a single Langmuir wave with frequency $\sim \omega_{pe,0}$ (Fig.\ref{expar} c). Simultaneously, both whistler and kinetic KAW are generated, which were investigated in a previous study\cite{che14prl}. The whistler wave dispersion relation $\omega=v_A^2kk_{\parallel}/\Omega_{ci}+\Omega_{ci}$\cite{gary93book} indicates that the whistler wave has ES component with frequency $\sim 0.001 \omega_{pe,0}$ and is strongly affected by density (Fig.\ref{disp} a, b). In Fig.\ref{disp} (c,d), we show both the electron and ion density fluctuations in $(\omega, k)$ phase space and find that both agree with the dispersion relation of ES whistler waves ($W_s$). Plasma fundamental emission  (Fig.\ref{eypar} c) is produced through both coalescence $L+W_s\rightarrow T$ and decay $L\rightarrow T + W_s$, where $T$ is the transverse emission. A new Langmuir wave $L^{\prime}$ is produced through $L+W_s\rightleftharpoons L^{\prime}$. The coalescence $L+L^{\prime}\rightleftharpoons T$ is much weaker because it is a second order process. As a result, harmonic emission in this stage is not identifiable. 
 
The ETSI reaches nonlinear saturation around $\omega_{pe,0}t \sim 10000$. The wavelength of Langmuir wave and ion caviton becomes longer and EM emission is produced in a broad range of frequencies and wave-numbers (panel d in Figs.~\ref{expar} \& \ref{eypar}), which is a consequence of repeating LC maintained by the feedback loop shown in Fig.\ref{chart}(see supplementary movie II). The turbulent fluctuations of density and electric field in phase space increase to levels comparable to that of the Langmuir waves and emissions. The emission reaches its balance between the plasmons of Langmuir waves and whistler waves --- the Manley-Rowe relation, and the coupling becomes $L+W_s \rightleftharpoons T$\cite{mel80ssr,diamond14book}. During this stage, electrons are strongly heated and scattered, the initial anisotropic electron beams become an isotropic halo population superposed over the core electron distribution function \cite{che14apjl}.

\section{L-W$_{s}$ Coupling and Langmuir Collapse }
\label{whistler}
The ES component of whistler wave, i.e. the ES whistler wave with frequency of several $\Omega_{ci}$ and wavelength $k\lambda_{De}<<1$ defines a slow time scale and a large spatial scale, while the Langmuir wave defines a fast time scale and small spatial scale $k\lambda_{De} \sim 1$. The coupling between ES whistler waves and Langmuir waves drives modulational instability that leads to the formation of long wavelength Langmuir envelopes $\mathbf{E}_L(\mathbf{x}, t)$ and cavitons. Langmuir waves exert a low frequency ponderomotive force on the motion of electrons and ions and mediate their interaction with whistler waves in a manner similar to L-IAW coupling\cite{zak72sjetp,rud78physrep}. The difference here is that 
whistler waves are produced in magnetized plasmas while IAWs are less sensitive to magnetic fields. The ES whistler wave is associated with EM whistler wave and cannot independently exist. In the following, we show this difference and why it does not significantly affect the critical condition for LC.

Assuming perturbations $\mathbf{E}=\mathbf{E}_s+ \mathbf{E}_L$, $\mathbf{v}_e=\mathbf{v}_s + \mathbf{v}_L$, $n_e=n_0+\delta n_L + \delta n_s$ and $\delta n_i\approx \delta n_s$, where the subscripts $L$ and $s$ represent fast and slow time scales, respectively. Neglecting of the high frequency interactions, we have the same driven equation as for the L-IAW coupling\cite{zak72sjetp}:
\begin{equation}
\frac{i}{\omega_{pe,0}}\frac{\partial \mathbf{E}_L}{\partial t} + \frac{\gamma v_{te}^2}{4\omega_{pe,0}^2}\triangledown^2 \mathbf{E}_L-\frac{\delta n_s}{2n_0}\mathbf{E}_L=0,
\label{fast} 
\end{equation}

On the slow time scale, ions play the same role as electrons in maintaining the cavitons. In Fig.\ref{disp}, both the ion and electron density fluctuations propagate at the phase speed of whistler waves $\sim 3v_A\sim 6 v_{ti}\sim 0.2 v_{te}$. The slow component of electron and ion motions in a magnetized plasma can be described by the following equations:
\begin{gather}
\label{ele}\nonumber\frac{\partial\mathbf{v}_s}{\partial t} +(\mathbf{v}_s\cdot \nabla)\mathbf{v}_s+\frac{e}{m_e}\mathbf{E}_s +\frac{e}{m_e c} \mathbf{v}_s\times \mathbf{B}_0\\
+\frac{\gamma_e T_e}{n_0 m_e}\nabla\delta n_s +\frac{1}{m_e n_0}\nabla\phi_{\rm{pm}}=0,\\
\label{ion} \nonumber\frac{\partial\mathbf{v}_i}{\partial t} +(\mathbf{v}_i\cdot \nabla)\mathbf{v}_i-\frac{e}{m_i}\mathbf{E}_s -\frac{e}{m_i c} \mathbf{v}_i\times \mathbf{B}_0\\
+\frac{\gamma_e T_i}{n_0 m_i}\nabla\delta n_s +\frac{m_e}{m_i^2n_0}\nabla\phi_{\rm{pm}}=0.
\end{gather}
where $\phi_{\rm{pm}}\equiv \vert E_L\vert^2/16\pi$ and $\nabla\phi_{\rm{pm}}$ is the ponderomotive force.

Eliminating $\mathbf{E}_s$ and using the approximation $m_e \mathbf{v}_s/m_i + \mathbf{v}_i\sim \mathbf{v}$, together with the ion continuity equation, $\partial (m_e+m_i)\delta n_s/\partial t +n_0(m_e+m_i)\nabla\cdot \mathbf{v})=0$, we obtain:
\begin{equation}
\frac{\partial^2}{\partial t^2}\frac{\delta n_s}{n_0} - c_s^2\nabla^2\frac{\delta n_s}{n_0}-M=\nabla^2\frac{\phi_{\rm{pm}}}{ n_0 m_i},
\label{slow}
\end{equation}
where $M=\mathbf{B}_0\cdot \nabla\times \mathbf{j}/(m_i n_0 c)$ is the modulation of the magnetic field that excites the whistler wave, $\mathbf{j}=en_0(v_i-v_s)$ is the current density and $c_s^2\approx(\gamma_e T_e+\gamma_i T_i)/m_i $ is the phase speed of the IAW. In a homogeneous plasma, to first order, the current density is caused by the polarization drift, i.e., $\mathbf{j}\approx~((m_e+m_i)~n_0/B_0^2) \partial \mathbf{E}_s/\partial t$. The curl in $M$ implies that ES whistler waves originate from the perpendicular EM components of whistler waves and KAWs\cite{stix92book}. In other words, the density fluctuations on the slow time scale are mediated predominantly by the EM whistler and KAW waves and the influence of $M$ is small. We will neglect $M$ when discussing the modulational instability and the critical condition of LC.

From Eq.~(\ref{fast}) and (\ref{slow}), the maximum growth rate for modulational instability is $\gamma_{m}=\omega_{pi}(\langle E_L^2\rangle/8\pi n_0 T_e)^{1/2}$ and the critical condition for LC is\cite{zak72sjetp}:
\begin{equation}
\frac{E_L^2}{8\pi n_e T_e}>\frac{1}{4}k_x^2\lambda_{De}^2.
\label{critic}
\end{equation}

In the nonlinear stage, the time scale of the modulational instability becomes longer than that it is in linear saturation due to the decrease of the electric field, but $(E^2/8\pi n_e T_e)^{1/2}\sim 0.2$ is still larger than the typical Langmuir wave-number $k\lambda_{De}/2\sim 0.05$ for the ES whistler waves, indicating LC can repeatedly occur. 
\section{Regeneration of Langmuir Waves}
\label{repeat}
LC transfers energy from large to small scales, inverse to the modulational instability. Repeating LC requires regeneration of Langmuir waves so that $L-W_s$ coupling can continue to produce emission (Fig.\ref{chart}).

In Eq.~(\ref{slow}), ions fill the cavitons and excite short wavelength IAWs \cite{galeev76ru} when the balance between the thermal pressure and radiation pressure is lost and Langmuir envelopes collapse. The dispersion relation of IAWs with thermal correction under the condition $v_{ti}<w/k<v_{te}$ is:
\begin{equation}
1+\frac{1}{k^2\lambda_{De}^2}-\frac{\omega_{pi}^2}{\omega^2}\left(1+\frac{3k^2v_{ti}^2}{\omega^2}\right)=0,
\end{equation}
where the term $3k^2v_{ti}^2/\omega^2$ in the bracket comes from the first order expansion of the ion zeta function $Z(w/k/2v_{ti})$, a higher order correction for the case $w/k/v_{ti}>1$, but not $\gg 1$. 

For short wavelength IAWs with $k\lambda_{De}\sim 1$ in a plasma with $T_i\sim T_e$, the dispersion relation becomes:
\begin{equation}
\omega^2\approx \omega^2_{pi}\frac{1+\sqrt{3T_i/T_e}}{4} + \sqrt{\frac{3T_e}{T_i}}k^2 v_{ti}^2,
\label{ia}
\end{equation} 
and for long wavelength $k\lambda_{De}<<1$,
\begin{equation}
\omega=\pm \sqrt{\frac{T_e}{m_i}} k.
\label{lia}
\end{equation}

Eq. (\ref{ia}) shows that the phase speed of short wavelength IAWs satisfies $\omega^2/k^2/v_{ti}^2\sim 1/k^2 \lambda_{De}^2$, and thus for the short wavelength IAW with a few tenth $k\lambda_{De}$, the exponential ion damping rate is comparable to electrons and the rate is $\gamma_{ia}\approx (m_e/m_i)^{1/2}\omega_{pi} /k^3\lambda_{De}^3$. The dissipation of the short wavelength IAW is slower by a factor of $\gamma_{ia}/\gamma_m\sim (m_e/m_i)^{1/2}$ than the modulational instability and thus this wave can be observed (Fig.\ref{expar} (b, c, d)). On the other hand, the damping rate of long wavelength IAW is too strong to maintain L-IAW coupling and is suppressed by L-Ws coupling.

During LC the wave energy is transferred from long wavelength Langmuir waves to short wavelength IAWs and then is returned to the newly generated short wavelength Langmuir waves. Such energy transfer can be shown in the phase space $(\omega, k)$ using quasi-particle (plasmon) description. The plasmon number is defined as $N=E_k/\omega_k$, where $E_k$ is the energy density of Langmuir envelope and $N(k,x,t)$ is the number of plasmons\cite{diamond14book}. The evolution of the mean plasmon number $\langle N(k,x,t)\rangle$ in phase space during wave interactions is determined by the wave self-interactions and wave-wave interactions (the details will be presented in a later paper):
\begin{gather}
\frac{d\langle E_k\rangle}{dt}=-\mathbf{v}_g\cdot \mathbf{D}\cdot \frac{\partial \langle N \rangle}{\partial \mathbf{k}},
\end{gather}
where the group velocity $\mathbf{v}_g=\mathbf{\partial \omega /\partial k}$, phase space diffusion coefficient $\mathbf{D}=\mathbf{q}\mathbf{q}\int\int_{\mathbf{q},\Omega} i\omega_{pe,0}^2\delta n_s^2 /(\Omega-\mathbf{q}\cdot \mathbf{v}_g+i\widehat{\Gamma}) d\mathbf{q}d\Omega$, $\Gamma$ is the self-interaction of Langmuir wave, the primary part $\langle\Gamma\rangle$ is associated with the mean plasmon number $\langle N \rangle$ and $\widehat{\Gamma}$ is the 
first order self-nonlinearity of Langmuir wave, such as linear growth or damping, where            $\mathbf{q}$ and $\Omega$ are the wave vector and frequency of IAWs, respectively.

In the case the linear growth $\widehat{\Gamma}\sim 0$, a second instability can occur at $\Omega \simeq \mathbf{q}\cdot \mathbf{v}_g$. If $\mathbf{q}\cdot\partial \langle N \rangle/\partial \mathbf{k}\vert_{\Omega}>0$, we have $d E_k/dt<0$, indicating the energy transfers from the Langmuir wave to IAWs. The Langmuir wave is depleted by the Landau damping of caviton trapped wave-scattering and the short wavelength IAWs repopulate the energy of short wavelength part of the energy distribution. If $\mathbf{q}\cdot\partial \langle N \rangle/\partial \mathbf{k}\vert_{\Omega}<0$, we have $d E_k/dt>0$, indicating the regeneration of Langmuir waves. The short wavelength IAW is damped by electrons and the hot electrons reproduce the short wavelength Langmuir waves.

The short wavelength IAW acts as an intermediary wave in the regeneration of Langmuir waves. The Langmuir wave energy gain by modulational instability and loss by LC can reach a balance, i.e. $\gamma_{ia} W_{ia}=\gamma_m W_L$, where the short wavelength IAW wave energy density $W_{ia}=n_0 T_e \sum_k \delta n_{s,k}^2 /n_0$, the short wavelength Langmuir wave energy density $W_L=\langle E_L^2\rangle(k_0/k)^{3/2}/8\pi$, and $k_0=1/\lambda_{De} (\langle E_L^2\rangle/ 8\pi T_e)^{1/2}$ --- the critical Langmuir wavenumber corresponding to LC at which the short wavelength IAW energy is transferred into short wavelength Langmuir waves\cite{galeev76ru}. 

The electron resonance with the short wavelength IAWs re-excites Langmuir waves with a frequency shift and Eq.~(\ref{fast}), when modified to include the short wavelength IAW excitation\cite{galeev76ru}, becomes:
\begin{gather}
\nonumber \frac{i}{\omega_{pe,0}}\frac{\partial \mathbf{E}_L}{\partial t} + \frac{\gamma v_{te}^2}{4\omega_{pe,0}^2}\triangledown^2  \mathbf{E_L}-\frac{\delta n_s}{2 n_0} \mathbf{E}_L=\frac{\delta n_{s,new}}{2n_0} \mathbf{E}_L \\
=-\frac{\omega_{pe}}{12} \sum_k \frac{\vert\delta n_{s}^k\vert^2}{n_0^2 k^2\lambda_{De}^2}\left(1+i\frac{2}{3}\frac{\gamma_k}{\omega_{pe}k^2\lambda_{De}^2}\right) \mathbf{E}_L, 
\label{newfast}
\end{gather}
where $\gamma_{k}$ is the damping rate of short wavelength Langmuir wave with wave number $k$. The first term on the right hand side of Eq.~(\ref{newfast}) is the frequency shift of a plain Langmuir wave by the scattering of the ion density fluctuations driven by short wavelength IAWs. The second term corresponds to the damping of long wavelength Langmuir waves due to their conversion to short wavelength IAWs. The frequency shift is
\begin{equation}
\delta \omega=-\frac{\omega_{pe}}{12}\sum_k \frac{\vert\delta n_s^k\vert^2}{n_0^2 k^2\lambda_{De}^2}.
\end{equation} 
This shift is the same for the entire Langmuir wavepacket spectrum and has no influence on the modulational instability.

After each LC, the frequency of the new Langmuir wave will decrease by a shift $\sim \omega_{pe} n_s^2/n_0^2/12\sim 0.01 \omega_{pe}$ assuming $k\lambda_{De}\sim 1$. We assume that the interval for LC is comparable to the time scale of modulational instability $100\omega_{pe}^{-1}$, the whole simulation is about 10000 $\omega_{pe}^{-1}$. Thus the total frequency shift is about $\omega_{pe}$. This agrees with what is shown in Fig~\ref{expar} in which Langmuir wave finally shifts to $k\lambda_{De}<<1$.
\section{Concluding Remarks}
PIC simulations were conducted to explore how the evolution of the strong ETSI produces Langmuir waves and plasma coherent emission. We found that LC plays a critical role in the process, which enables regeneration of Langmuir waves and maintains a feedback loop for emission beyond the linear ETSI saturation (Fig.\ref{chart}). The onset of LC is introduced by the L-L wave coupling at the ETSI linear saturation stage and maintained by L-Ws coupling from the nonlinear decay stage to the nonlinear saturation. The low frequency KAWs and whistler waves generated near ETSI peak finally reach equilibrium with the non-Maxwellian electron velocity distribution function (e.g. core-halo structure), as found in previous studies\cite{che14prl,che14apjl}.

In our simulations, the ETSI nonlinear saturation time is $\sim 1.5\times 10^4\omega_{pe}^{-1}$. Because the modulational instability nearly dominates the entire process, the nonlinear saturation time is approximately proportional to $(m_i/m_e)^{1/2}$, and for the physical mass ratio, the ETSI nonlinear saturation time should be $\sim 10^5 \omega_{pe}^{-1}$, which is significantly longer than the ETSI linear saturation time $(n_0/n_b)^{1/3} \omega_{pe}^{-1}\sim 2 \omega_{pe}^{-1}$. Note that our simulation assumes instantaneous injection of the electron beam, while in the corona the electron-acceleration time is finite and the beam will propagate out of the region of initial generation. The acceleration time also affects the actual duration of the bursts\cite{goldstein79apj,rat14aap}. The overall scenario is that coronal bursts produce non-thermal electrons that escape into space and produce interplanetary bursts\cite{aschw02ssr} with accompanying waves. Our simulation assumes the beam energy is about 100 times that of the coronal thermal energy. For nanoflares, the beam energy is about keV if the corona temperature is $\sim10$ eV. For flares, the electron beam energy can reach MeV. The larger beam energy will change the results slightly since the ETSI growth rate does not rely on the velocity once the threshold is reached, but the turbulence becomes stronger and the decay lasts longer. On the other hand, we can estimate the emission power from Fig.\ref{eypar} (b, c, d) in which the intensity ratio of the $E_y$ of the radiation and the Langmuir wave is about 0.01-- 0.001, and thus the emission power is about a factor of $ 10^{-4}-10^{-6}$ of the Langmuir wave power. Such energy loss is negligible dynamically. Therefore, the mechanism we have explored can provide a complete and self-consistent solution to the long-standing puzzle of why the duration of solar radio bursts is much longer than the linear saturation time of the ETSI (``Sturrock's dilemma"\cite{sturrock64}). 

The short wavelength IAWs and ion cavitons are two characteristics of LC, and can be detected by in-situ solar wind observations. In particular, the forthcoming Solar Probe Plus mission will be capable of in-situ detection of such radiation at 10 $R_{\odot}$ from the Sun. The newly launched Magnetospheric Multiscale Mission is capable of in-situ detections of Langmuir waves and cavitons in magnetosphere and solar wind at 1AU.

We summarize some basic observations which are consistent with our model predictions in Table 1. 
\appendix
\section*{ L-L Coupling and Emission}
The Langmuir waves generated in solitary wave trapped electrons $L_l$ propagate forward while the Langmuir waves generated in background electrons $L_h$ propagate backward. In the following we will clarify how the two Langmuir waves propagating in opposite direction produce emission through $L_h+L_l\rightarrow T$. 

\begin{gather}
\label{disper1}\omega_{L}\approx \omega_{pe} +3k_L^2 v_{te}^2/2\omega_{pe}, \\
\label{disper2}\omega_{T}\approx (\omega_{pe}^2 +k_T^2 c^2)^{1/2},
\end{gather}
where the dispersion relation of Langmuir wave is approximated under the condition $k\lambda_{De,0} << 1$.

\begin{gather}
\omega_{h}\pm\omega_{l}=\omega_{T} \\
\mathbf{k_h} + \mathbf{k_l} =\mathbf{k_T}.
\end{gather}
Fig.\ref{eypar}(b) shows that both the fundamental and harmonic emissions have $k_{T}\lambda_{De,0} <v_{te}/c\sim 0.1$, thus we approximate the dispersion relation of the emission as $\omega_{T}\approx \omega_{pe} +k_T^2 c^2/2\omega_{pe}$. It is easy to show that the plus sign in the selection rule requires $\mathbf{k_h}\mathbf{k_l}<0$, i.e. the two Langmuir waves must be anti-parallel. The resulting $\mathbf{k_{T}}$ is more likely to be positive and the emission propagates forward with harmonic frequency $\omega/\omega_{pe}\sim 1.6$.
\section{Acknowledgements}
HC and PHD thank participants for discussions in the``8th Festival de Th\'eorie", Aix-en-Provence, France, 2015. HC is supported by the NASA Magnetospheric Multiscale Mission in association with NASA contract NNG04EB99C. PHD thanks M. Malkov for discussions and the DOE grant No. DE-FG02-04ER54738 for support. The simulations and analysis were carried out at the NASA Advanced Supercomputing (NAS) facility at Ames Research Center under NASA High-End Computing Program awards SMD-14-4848 and SMD-15-5715. 

\end{article}

\begin{table}
\caption{Model Predictions and Observational Evidence}
\renewcommand{\arraystretch}{1.2}
 \begin{tabular}{c | c | c}
 \hline\hline 
 Model Predictions & Observations & References \\
 \hline
 In the solar corona emission duration  & Coronal Type J \& U radio bursts, & \cite{aschw95apj,aschw02ssr,saint13apj}  \\
 $\sim 10^5 \omega_{pe}^{-1} \sim 1-10$~ms. &
  Weak Coronal Type III radio bursts. &   \\
 \hline
 Langmuir waves \& whistler waves & Interplanetary Type III radio bursts & \cite{kellogg92grlb,mac96aa,ergun08prl} \\
 \hline
 Langmuire collapse \& short wavelength IAW & Interplanetary Type III radio bursts & \cite{lin81apj,kellogg92grla,theja04npg}\\
 \hline
\end{tabular}
\end{table}

\end{document}